# Commentary: The nature of cancer research

Steven A. Frank

Department of Ecology and Evolutionary Biology, University of California, Irvine, CA

92697-2525, USA. E-mail: safrank@uci.edu

June 16, 2015  15:36

Cancer research reflects an implicit conflict. On the one hand, there is an overwhelming desire to control the disease. We all wish that. On the other hand, we would like to understand why cancer follows so many clearly defined yet puzzling patterns. Why is there such regularity in the rates of progression? Why do different tissues vary so much?

There should, of course, be no conflict between control and understanding. But the history of cancer research seems to say that those different goals remain oddly estranged. Peto's 1977 article[1] locates the seeds of this conflict most clearly. He describes what is still the most powerful theoretical perspective for analyzing the causes of cancer. He presents many key unsolved puzzles within that context. He also says why most cancer researchers are not interested in these fundamental issues.

The subsequent decades of research grew around this rift, blindly, in the way that research disciplines often grow. Let us revisit Peto, almost 40 years ago. We can learn much about the current nature of cancer research.

## The rift in cancer research

What Peto does most profoundly is to tell us what we do not understand about cancer. He does this by describing unsolved puzzles. Peto's puzzles remain the best framing for what it would mean to understand cancer. Before turning to those puzzles, let us consider how Peto places the particular biological problems within the context of cancer research in the 1970s.

Peto begins by noting the strange rift between what most cancer researchers think about and the deep unsolved puzzles of cancer. His first paragraph includes

> It seems likely that multistage processes underlie the generation of a large majority of human cancers … yet most research workers do not have any real interest in discussing the various alternative multistage models that attempt to describe these processes. These research

> workers can't all be wrong, so what is wrong with multistage models? The trouble is, I
> suppose, that the processes usually invoked are, in principle, extremely difficult to observe
> … and that very similar predictions for the few things we can actually observe may follow
> from mathematical elaboration of various very different multistage models …

Good reasons for a practical person to ignore the broader framing of multistage processes.
However, Peto sees potential for deeper understanding and the benefits that such understanding
will bring

> No single process is likely to be the whole truth, and we must hope that some grand
> synthesis of the known processes will eventually be put together which will describe all the
> essential features of human cancer induction. Although the eventual synthesis is not yet in
> sight, multistage models should at present be thought about to some extent, and their general
> features should be common knowledge, as the general framework of this eventual synthesis
> will … almost certainly be some kind of multistage model.

Followed by a hint of puzzles to come

> Moreover, despite all their present uncertainties, multistage models for carcinoma induction
> have already offered plausible answers to various questions concerning monoclonality, dose-
> response relationships under conditions of regular exposure, hypothetical "threshold" doses,
> the synergistic effects of different carcinogens, the role of luck, and, last but not least, the
> connection between cancer and aging.

Peto felt strongly about the great insight that might come from asking the right questions. Yet he
never forgot that he was running against the crowd. The first two sentences of the article are

> This section is intended as an introduction to, and apology for, multistage models. Such
> models occupy a curious position in the world of cancer research.

That conflict is perhaps what gives his article an unusual sense of clarity—of seeing both the
great challenges of scientific understanding and the inherent limitations of scientific research.
Peto says no more about those philosophical issues. So let us follow him to the fascinating
biological puzzles of cancer. And, perhaps more importantly, let us consider what we can learn
about the nature of cancer research. (See my monograph[2] for detailed analysis of the issues
discussed in this commentary.)

## Why do tissues vary?

Some tissues suffer high rates of cancer, such as breast, prostate, colon, and bladder. Other tissues have lower rates, such as kidney, brain, and liver. Some tissues have cancers mostly early in life, such as retina and bone. Other tissues have cancers mostly later in life, such as lung and pancreas.

The clarity of these patterns sets a challenge. Any conceptual understanding of cancer must explain these tissue-specific variations.

Many of Peto's specific puzzles build on this basal challenge. His hypothesis is simple. Cancer is a disease of cell division. The cells of epithelial tissues often divide throughout life, whereas the cells of other tissues typically divide less. That continual renewal of epithelial tissues means a lot of cell division, and thus a lot of cancer. Risk often goes up with a high exponent of the number of cell divisions. Thus, most cancers are late-life carcinomas of epithelial tissues. By contrast, the retina divides early in life and then mostly stops. Retinoblastoma is rare, and when it occurs, it mostly happens before age five. In bone tissue, much cell division happens during the growth spurt of the teenage years, during which most osteosarcomas occur.

Cell division depends on tissue architecture. In continually renewing epithelial tissues, most cell division occurs in short-lived cell lineages that die out quickly, with little chance of accumulating changes that lead to cancer. The tissue renews from long-lived stem cell lineages that divide rarely. To understand the risk and the timing of cancer, one must understand the nature of cellular renewal. That renewal must be understood in the context of the multiple protections against cancer that ultimately break down before disease progresses.

The outlines of an explanation take shape. The architecture and renewal of a tissue determine the number of divisions in long-lived lineages. The different protections against cancer set the multiple barriers that must be overcome before disease occurs. Peto's final paragraph emphasizes that stochasticity, or luck, plays an essential role

> To reach any useful conclusions from these fundamental hypotheses, one has to add extra assumptions … However, all have in common a stochastic approach rather than a deterministic one. Put another way … luck has an *essential* role … in determining who gets cancer and who does not. The probability of one of my bronchial cells generating a fatal carcinoma can be predicted, but whether in actual fact one cell will do so cannot.

## Problems of history

Peto credits Cairns' 1975 article[3] for the key ideas about cell division and tissue architecture. I do not know of any significant antecedents to Cairns' work. However, close historical research typically uncovers many precursors. Cairns acknowledged Christopher Potten, whose work on stem cells and tissue renewal[4] certainly contributed important data and a conceptual frame for building an explanation. Very likely, others had developed ideas about cell division in relation to differences in cancer incidence between tissues. But the origins of these ideas remain shrouded in a rarely studied history.

The roles of history and of ideas in cancer research are peculiar. To overstate a bit, it seems as if each investigator or subdiscipline views its own recent work as being the first to crack open the hard nut of cancer causality and control. I have often wondered whether there is something unusual about cancer that causes cancer biologists to be even less interested than usual in alternative perspectives and in the history of ideas. Here is an example.

Tomasetti and Vogelstein's recent article[5] on the roles of cell division and luck in cancer risk attracted much attention. The argument is the same as Peto's, almost to the word. From Tomasetti and Vogelstein's introduction

> If hereditary and environmental factors cannot fully explain the differences in organ-specific cancer risk, how else can these differences be explained? Here, we consider a third factor: the stochastic effects associated with the lifetime number of stem cell divisions within each tissue.

I reviewed this manuscript for the journal *Science* before it was accepted for publication. I recommended publication, because I valued the clear summary of recent data relating stem cell divisions to cancer incidence. But I also noted that

> I am a bit shocked by the introduction and the framing of the argument. I guess the standard in molecular biology is that to think of an idea is equivalent to being the first person to have ever thought of the idea. And maybe it does not really matter.

It does not matter because, as Peto's article taught me, cancer research is a discipline that lacks interest in ideas and history. Fighting against the strong current of "new" work leads only to loss of energy. But this episode also made me wonder once again about the nature of cancer research.

## The nature of cancer research

In particular, Vogelstein is, to me, one of the truly great heroes of modern cancer research. No one has done more to sharpen our focus on the multistage character of cancer progression. But why is it that Vogelstein is the key figure of the multistage perspective, based on his work starting in the late 1980s? Peto's 1977 article is all about the essential power of the multistage perspective for making sense of so many aspects of cancer biology. And Peto's article was itself the culmination of decades of theory, going back to Armitage and Doll[6], and decades of deeply insightful empirical work on genetics[7], including the early work of Ashley[8] and Knudson[9].

Vogelstein is the modern hero because he connected genetics to the mechanisms of change in cells and tissues[10]. This connection opened the way for using the power of genetic technology to get at the biochemical and regulatory changes that happen during cancer progression. It is reasonable to think that only through mechanism can we achieve both an understanding of causality and the potential for control. What came before, including Peto's article, mattered very little as the new prospects for molecular study opened up.

I think that crude summary is mostly right about the history and about cancer research. The new molecular technologies taught us so much, so fast, that it hardly mattered what came before. But, inevitably, the need to think through what all of the data really mean has once again become a limiting factor. Indeed, the easier it is to obtain great amounts of data, the more one needs to frame the issues in a meaningful way[11]. To get the true modern value of Peto's way of thinking about particular biological problems of cancer, one must first come to some explicit understanding of the intrinsic rift in cancer research that caused Peto to apologize for his approach in his first paragraph. I repeat here the key lines quoted above

It seems likely that multistage processes underlie the generation of a large majority of human cancers … yet most research workers do not have any real interest in discussing the various alternative multistage models that attempt to describe these processes. These research workers can't all be wrong, so what is wrong with multistage models? The trouble is, I suppose, that the processes usually invoked are, in principle, extremely difficult to observe … and that very similar predictions for the few things we can actually observe may follow from mathematical elaboration of various very different multistage models …

The next section considers the oddness of cancer research from one additional perspective. After that, we will be prepared to consider Peto's discussion of particular puzzles of cancer in relation to modern problems of cancer research.

## Control versus causality and the nonintuitive scale of risk

I consider Peto's concerns about cancer research to be real, and to pose an unanswered question. Superficially, it is easy enough to say that understanding mechanism is what really matters for making progress on cancer, and so is what really matters to most cancer biologists. By that view, ideas about multistage causality and about the history of ideas belong to mathematics and to philosophy. Only empirical proof of mechanism matters for biology.

However, I think there is a special aspect of cancer that inevitably distorts perspectives on what is truly a key issue. The multistage nature of cancer that meant so much to Peto has a special consequence for understanding risk factors and for thinking about control.

Consider a simple example, in which $n$ independent events must happen before cancer develops. Suppose, for example, that one of the events is the escape, by a clone of rapidly dividing cells, from suppression by the immune system. From the perspective of preventing cancer or controlling disease, suppression of tumors by immunity is sufficient by itself. With suppression by immunity, no disease. By contrast, all aggressive tumors associate with lack of suppression by immunity. From the perspective of causality, enhancing suppression by immunity greatly reduces disease, whereas blocking suppression by immunity greatly increases disease. This single factor of immunity seems, by itself, all we need to focus on with respect to both control and causality.

The other $n - 1$ factors required for cancer are equally powerful for control or for explaining causation. If we focus solely on any single factor, each seems to be nearly sufficient by itself. Of course, everyone knows that cancer requires many changes, and that a single factor alone is not sufficient. Nonetheless, cancer research seemingly fractures into nearly independent subdisciplines, each with its own focal mechanism that holds out the greatest promise for cure and that holds a special place in the causal scheme. That distortion of perspective arises from the true underlying multistage nature of cancer. Multistage causality distorts, in a nonintuitive way, the relation between single mechanistic factors and an overall understanding of control and causality. That is, in my view, one of Peto's key messages, and also why he knew that his messages would not be heard. He begins the final section of his article with

> If a synthesis of several different lines of evidence relating to the mechanisms of cancer is to emerge, then, at least for carcinomas, it seems that such a synthesis will only be achieved in the framework of one or another of the possible multistage models. All of these models indicate that the final incidence rate of cancer is the arithmetic product of more than one term, the different terms each being dependent on different causative agents or processes. If this is so, it may be misleading to ask what *the* cause of a certain type of cancer is, as if there was one fundamental cause and all else was ancillary. For example, if environmental mutagens can be identified and manipulated to halve the net mutation rate in a certain tissue, then the cancer rate could be correspondingly reduced, but an equivalent improvement might be equally achievable by manipulating the environmental determinants of some other, qualitatively different, necessary cause in the sequence of changes which culminates in malignancy. It is still an open question, to be answered separately for each different type of cancer, as to which class of causes can most easily be identified and reduced, even if, thanks to recent improvements in methods for determining mutagenicity, the most rapid progress in determining causes over the next few years results from the study of mutagens.

Currently, instead of mutagens, we may substitute the study of gene mutations, or epigenetic changes, or gene network perturbations, or immunity. All of those factors are important. But to understand *cause* in a meaningful way, each factor must be understood as part of a multistage process in which *rates* of change in individual factors affect, in potentially nonintuitive ways, the *rates* of cancer incidence at different ages. Testing hypotheses about individual causal factors can only be done within a framework of the various multistage rate processes that interact to determine the final outcome: cancer or no cancer at each age.

## Peto's puzzles of cancer

This section lists the interesting ideas raised by Peto. The following section analyzes the most important idea for moving current research ahead. That important idea concerns the clear but unexplained pattern by which cancer incidence changes in relation to the dosage, timing, and duration of carcinogen exposure.

Some of the following ideas may be wrong. But even the wrong ideas fail in interesting ways. The value, as Peto noted, was in the questions

> Because I am trying to illustrate how one might, when considering new ideas, do so in the context of multistage models, I have introduced some ideas … which eventually may be found to be false …

**_Multistage progression and the right way to understand the causes of cancer._** Peto organizes his many puzzles of cancer around multistage progression. The idea of multistage progression is not itself directly testable. Essentially any pattern can fit within it. Instead, multistage models are tools by which one develops a useful comparative prediction. Comparative predictions follow the structure of this example: If a genetic change or a carcinogenic exposure influences an early stage in cancer progression, then such a change or exposure early in life will have a stronger effect on cancer incidence than such a change or exposure late in life.

The key point concerns what it means to say that a factor has a causal influence on cancer[2]. Currently, the most common approach compares genetically engineered mice or other animals. If a gene _causes_ cancer, then comparing animals with and without a change in that gene should change the incidence of tumors. A change in incidence is usually measured by a significant shift in the overall incidence pattern of tumors, typically by comparison of Kaplan-Meier curves.

There is nothing inherently wrong with such comparative genetic tests. The problem, in the context of Peto's vision, is that the way in which such comparative genetic tests are formulated and analyzed cannot lead to any truly meaningful notion of causality. A meaningful notion of causality tells us how changes in genes or in carcinogenic exposures alter the interaction of multistage rate processes. Those interactions between multistage rate processes together determine the shift in the risk of cancer at each age in response to the changes in the hypothesized causal factor.

That "meaningful notion of causality" is, of course, harder to achieve that a simple genetic manipulation and comparative Kaplan-Meier plot of tumor incidence. But modern technologies combined with both the older carcinogenic tools and Peto's vision for testing hypotheses of causality could achieve the more profound analysis of cause. Such progress would require joining modern work with an older line of thought that has mostly been lost. The following puzzles from Peto recall that older view.

***What explains differences between tissues?*** Multistage theory makes a simple comparative prediction. More cell division correlates with more cancer. Another comparative prediction: continually dividing tissues, such as lung, have cancers concentrated late in life, whereas early dividing tissues, such as the retina, have cancers concentrated early in life. Or, originally from Cairns[3], tissues with a stem hierarchy that reduces division in long-lived lineages are more protected against cancer than tissues with a more even distribution of division among cells.

***What explains departures from log-log linearity of incidence?*** The simplest multistage models predict a linear increase in incidence with age on a log-log plot. The incidence patterns for many epithelial cancers roughly follow log-log linearity, but also tend to depart from pure log-log linearity in characteristic ways. Why? Peto lists many possibilities: genetic heterogeneity, environmental heterogeneity, age-related changes in carcinogen exposure, cohort changes in exposure or behavior, diagnostic variations in reporting cases, and the saturating effects late in life when most individuals have progressed through the early stages. Here, we have many plausible explanations. So many, that the observed patterns are inexplicable without some additional leverage provided by a more incisive comparative prediction. Below, I return to the issue of leverage via comparative prediction.

***Do other noncancerous diseases also arise by multistage processes of local cellular changes and subsequent expansion or spread?*** Peto considers whether atherosclerotic plaques arise from local cellular changes followed by clonal expansion. If so, then perhaps the genesis of heart disease is conceptually similar to cancer, following a multistage process of breakdown in the protections against disease. Recently, I have been interested in whether neurodegenerative diseases may also arise initially by multistage changes in small local pieces of tissue, followed by spread from a disease focus[12]. Peto's point is that the multistage nature of cancer may be part of a wider perspective on the multistage nature of disease progression, particularly for those diseases that increase in incidence exponentially later in life.

***What are the relative roles of mutagenicity and mitosis?*** The small and large intestines include epithelial tissue that divides continuously throughout life. Yet cancers of the small intestine are rare, whereas cancers of the large intestine are common. Cell division by itself cannot explain all variations between tissues. Perhaps the large intestine suffers greater exposure to mutagens. But mutagenicity cannot be the sole additional factor needed to explain variation between tissues, because of the next puzzle.

***What explains differences between mice, humans, and whales?*** A human has 1000 times as many cells as a mouse and lives at least 30 times as long. If one uses the same parameters for multistage progression in mice and humans, then humans would have immensely greater cancer risk than mice, because humans have so many more cell divisions[13]. Yet the actual rates do not differ greatly. And why don't whales have much more cancer than humans?

One of the most important comments in Peto's article concerns how to approach such apparently inexplicable puzzles

> The most direct way to elucidate induction mechanisms for human carcinomas is to study the characteristics of epidemiologically determined causes of human carcinomas, and, in particular, we must look at what happens when a known cause is applied at a different dose rate or for a different time period.

The point is simply that the patterns by themselves will not reveal the underlying processes. Too many different processes lead to the same epidemiological patterns, all consistent with a wide variety of multistage models. To understand why humans get cancers at different rates from mice for a given number of cell divisions, we have to measure how changes in carcinogenic processes alter cancer incidence in humans and in mice. Presumably, humans are less sensitive to carcinogenic processes than are mice. In this context, cell division is itself understood to be an important carcinogenic processes. To understand the differences between humans and mice would mean to understand exactly how the different sensitivities to carcinogenic processes arise.

Note that the ultimate epidemiological pattern of incidence is the only thing that truly matters, because it is the only direct measure of actual cancer cases in relation to an underlying variable, such as cell division or species type. The epidemiological pattern reflects all of the underlying rate processes of cancer progression that interest us. We just have to learn how to read those patterns.

***Why does lung cancer incidence level off after cessation of smoking?*** Smokers have higher incidence of lung cancer than nonsmokers. The increase in incidence for smokers relative to nonsmokers continues to rise with the number of years of smoking. If a smoker quits, the excess incidence at the age of quitting continues throughout life, but does not change much in the years after quitting. Roughly speaking, the incidence rate levels off after cessation of smoking. Peto comments that this observation is "one of the strongest, and hence most useful, observational restrictions on the formulation of multistage models for lung cancer."

Peto argues that the data are consistent with a simple multistage model in which smoking increases the event rate of the penultimate stage in the development of cancer. The idea is that smoking moves individuals quickly through that penultimate stage to a waiting class just one stage before cancer. The quitters who have moved through that penultimate stage while smoking will have only the final stage remaining before suffering cancer. If they move through that final stage at a constant rate, then incidence per year will be approximately constant after quitting. This argument for smoking's effect on the penultimate stage of lung cancer progression has been repeated many times[2].

I think that, for mathematical reasons alone, this is a very weak argument. Many plausible alternative models also fit the data. In spite of the fact that the particular argument is perhaps not so strong, I believe this way of thinking about the relation between variations in carcinogen exposure and incidence does provide the best, and perhaps only, way to connect mechanism to the consequences for outcome—to connect the biochemical and cellular effects of a cause to the ultimate consequence for the appearance of cancer. The conceptual approach is right, but the particular example is not a good one. I return to these issues below.

***How do variations in timing and dosage of carcinogen exposure affect the patterns of incidence?*** Peto discusses changes in lung cancer incidence in relation to the number of cigarettes smoked per day (dosage) and the delay in the age at which smoking starts (timing). Consider the possible consequences of dosage in terms of multistage progression. If smoking affects only one stage among many in progression, then increasing dose saturates the change in incidence by altering a single normally limiting step into a process that proceeds at such a high rate that it is no longer limiting. At a saturating dose, the incidence curve will trace a higher rate of progression to disease but will have a lower slope with respect to age, because fewer limiting steps remain. In simple multistage theory, the log-log slope is the number of limiting steps minus one. The decrease in log-log slope caused by saturating carcinogen exposure should be roughly equal to the number of stages affected by the carcinogen.

With regard to timing, several aspects may be important. For example, if a carcinogen affects only an early stage in progression, then exposure early in life should have a stronger effect on incidence than exposure late in life. Duration of exposure can also be analyzed in relation to the number of stages affected and the order in which those stages typically limit the development of a tumor.

These ideas about timing, duration, and dosage provide a rich set of hypotheses that can be used to understand comparative studies of subjects with different exposures or to design experimental studies in which these factors are manipulated. For any type of cancer, the predicted response concerns changes in the age-incidence pattern with respect to changes in hypothesized factors, when evaluated in terms of an underlying causal scheme within a multistage perspective.  Peto discusses several patterns in the data available at that time, and how one might interpret those patterns within a multistage context. For example, Druckrey[14] summarized many different studies that showed an overall regularity of the responses to variations in carcinogenic dosage and duration. Many of those observations remain unsolved puzzles. Solutions would provide deep insight into the causes of cancer.

***Among carcinogens, how do the effects of mutagens and mitogens differ?*** Around the time of Peto's article, there was great interest in quantifying the mutagenic effects of chemical agents[15]. The argument was that the most important carcinogens with respect to risk are typically mutagenic. Peto places this argument about mutagens into the complex observations about how different kinds of chemical agents affect cancer incidence.

To simplify greatly, we might compare mutagens with mitogens. In theory, if one applies a mutagen to initiate cells with mutations, followed by a mitogen that promotes expansion of cellular clones carrying the mutation, then many target cells are present with the risk-bearing mutation, and the effect on incidence should be great. By contrast, if one applies a mitogen first, followed by a mutagen, then the initial clonal expansion does not raise the number of risk-bearing mutated cells, and the associated effect on incidence should be small.

Peto ties that classical two-stage model of initiators (mutagens) and promoters (mitogens) to a broader perspective on multistage progression. In the context of multistage progression, one may think about how timing, duration, and dosage of mutagens and mitogens may differ in their consequences for the observed age-incidence patterns. Many scattered and apparently unconnected observations could potentially be unified within this perspective.

## The most important lesson for current research

Peto concludes his summary of carcinogen exposure in relation to cancer incidence by highlighting what he considers to be one of the great unsolved puzzles: that dosage has a much weaker effect than duration of exposure

> The fact that the exponent of dose rate is so much lower than the exponent of time is one of the most important observations about the induction of carcinomas, and everyone should be familiar with it—and slightly puzzled by it!

Throughout his article, Peto develops a vision of how we should recognize such puzzles, why they are so important for understanding the causes of cancer, and how we should go about analyzing the causes of cancer. His vision is the multistage nature of cancer. Stages are understood as rate processes that define the change in some factor or barrier that must be altered in order for cancer ultimately to arise. Each stage, or cause, can only be understood in relation to how that particular rate process interacts with other processes to determine the overall rate at which cancer develops at each age. The age-incidence curve is the ultimate measure of outcome. *Cause* means the way in which a changed input, such as an altered gene or tissue or exposure, changes the output, a shift in the age-incidence curve.

The network of rate processes through which changed inputs shift the age-incidence pattern is too complex to understand intuitively. One must compare a set of clearly formulated hypotheses about the overall multistage process, and use those hypotheses to make specific predictions about the relations between altered inputs and shifts in age-incidence.

In 1977, that was a brilliantly complete vision of what must be done. But it was hard to put such a demanding approach into practice. Today, we have vastly greater opportunities provided by technology. The current technologies tend to focus on genetic or epigenetic aspects. We could expand the tools available by reconsidering the powerful earlier methods with respect to chemical exposures of mitogens, mutagens, and other carcinogenic factors. Perhaps we only need to connect the new technologies with the older, broader concept of carcinogens, interpreted within Peto's vision for how to study cause.

It should be possible to design experiments that alter genetics, epigenetics, and a variety of carcinogenic processes. A hypothetical scheme of multistage causal interactions should allow one to generate testable predictions about how particular combinations of changes would shift the age-incidence response. Causal mechanisms could be evaluated through iteration of multistage model, prediction about outcome, and experimental test[2]. Such a program, although ambitious, may be within reach. And, as Peto implied, there really is no other choice for understanding the causes of cancer.

## Conclusion

Peto's article expresses great biological insight and mature scientific wisdom. He wrote the article in his early 30s. He followed with one of the brilliant careers in cancer research. His primary accomplishments reflect, inevitably, what could be done on epidemiological correlates, dose-response relations, and clinical trials. He left behind the intriguing puzzles and grand synthesis of his 1977 article. The fact that Peto rarely looked back says a lot about the nature of cancer research.

Must it be so? In this era of immense technological power and opportunity for data collection at all scales, perhaps Peto's puzzles will return to inform a deeper understanding.